\documentclass[%
 pre,
 twocolumn,
superscriptaddress,
 amsmath,amssymb,
 aps,
floatfix,
]{revtex4-2}

\usepackage{graphicx}%
\usepackage{dcolumn}
\usepackage{bm,bbm}
\usepackage{xcolor}
\usepackage{xfrac}
\usepackage{amsmath,amssymb}
\usepackage{pifont}
\usepackage{array}
\usepackage{soul}

\usepackage{tabularray}

\usepackage{ulem}

\usepackage[colorlinks,linkcolor={blue},citecolor={blue}, urlcolor={blue}]{hyperref}
\definecolor{armygreen}{rgb}{0.55, 0.73, 0.0}
\definecolor{darkblue}{rgb}{0.0, 0.0, 0.41}

\usepackage{tikz}

\begin{document}

\title{An Orbifold Framework for Classifying Layer Groups \\ with an Application to Knitted Fabrics}

\author{Sonia Mahmoudi}
\email{sonia.mahmoudi@tohoku.ac.jp}
\affiliation{%
Advanced Institute for Materials Research (WPI-AIMR), Tohoku University, 2-1-1 Katahira, Aoba-ku, Sendai, Miyagi 980-8577, Japan
}
\affiliation{%
RIKEN, Center for Interdisciplinary Theoretical and Mathematical Sciences (iTHEMS), 2-1, Hirosawa, Wako, Saitama 351-0198, Japan
}
\affiliation{%
International Institute for Sustainability with Knotted Chiral Meta Matter (WPI-SKCM$^2$), Hiroshima University, 1-3-1 Kagamiyama, Higashi-Hiroshima, Hiroshima 739-8526, Japan
}

\author{Elizabeth J. Dresselhaus}
\affiliation{%
International Institute for Sustainability with Knotted Chiral Meta Matter (WPI-SKCM$^2$), Hiroshima University, 1-3-1 Kagamiyama, Higashi-Hiroshima, Hiroshima 739-8526, Japan
}
\affiliation{University of California, Berkeley, CA 94720, USA}

\author{Michael S. Dimitriyev}
\affiliation{%
Department of Materials Science and Engineering, Texas A\&M University, College Station, TX 77843, USA
}

\date{\today}

\begin{abstract}
Entangled structures such as textiles and architected materials are often doubly periodic. Due to this property and their finite transverse thickness,  the symmetries of these materials are described by the crystallographic layer groups. While orbifold notation provides a compact topological description and classification of the planar wallpaper groups, no analogous framework has been available for the spatial layer groups. In this article we develop an orbifold theory in three dimensions and introduce a complete set of Conway-type symbols for all layer groups. To illustrate its applicability, we analyze several knitted fabric motifs and show how their layer-group symmetries are naturally expressed in this new orbifold notation. This work establishes a foundation for the topological classification of doubly periodic structures beyond the planar setting.
\end{abstract}

\maketitle

\section{Introduction}
Today, researchers are rapidly expanding the range and application of entangled metamaterials, from engineering efforts towards programmed active textiles~\cite{sanchez2021_textileTechWearables} to weaving at the molecular scale~\cite{Liu2016_WeavingCOF}. Additive manufacturing enables novel textile designs~\cite{Keefe2022_TextileAM} and other three dimensional architected materials~\cite{Zhou2025_Polycatenated}. 
In order to build a generic framework for understanding and designing entangled metamaterials, we need new mathematical tools to concisely describe the symmetries and structures within these increasingly complex materials.

\begin{figure}[t]
    \centering
\includegraphics[width=\linewidth]{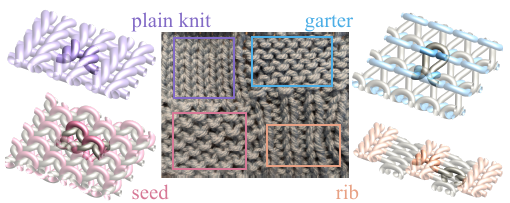}
    \caption{Fabric swatch with each quadrant a different basic knitted stitch along with a schematic of the stitch structure- upper left stockinette (as known as plain knit or jersey knit), upper right garter stitch, lower left seed stitch and lower right rib stitch.}
    \label{fig:four_stitches}
\end{figure}

The topology and symmetry of many of these materials are described not by the two-dimensional wallpaper groups but by the three-dimensional layer groups~\cite{ITE-layer,Sonia-Symmetry}. Such structures are periodic in two directions while possessing a finite thickness in the third. Their mechanical and geometric properties are strongly influenced by the symmetry constraints imposed by these subperiodic crystallographic groups~\cite{Liu2018_periodicKnots}. In the case of knitted fabric structures, the geometry is also determined by the necessity that yarn torsion is continuous except at frictional contacts between yarns \cite{leaf_glaskin_1955_knitted_loop} and the mechanics is also influenced by friction and yarn tension. In this work we consider the symmetries of idealized fabric structures arising from numerical simulations detailed in \cite{Singal2023}. 

The mathematical study of symmetry in fabrics was significantly advanced by Roth (1993), who analyzed isonemal two-way twofold woven structures~\cite{Roth}, and was more recently extended in~\cite{Penas2024} to include all layer-group symmetries arising in two-way twofold and three-way threefold weaves. Recent developments extended symmetry-based approaches to the study of 2-periodic polycatenane, knit, braid, weave, entangled graph, and crystal patterns~\cite{O'Keeffe2periodic,Evans-Hyde2022,Okeeffe2024,Fu_2024,takano2025}, but these studies did not incorporate orbifold methods. Existing orbifold techniques have been applied in the case of 3-periodic entangled structures, were orbifolds of the hyperbolic plane are muted to be compatible with symmetry of triply periodic minimal surfaces~\cite{Periodic1,Periodic2,Periodic3,Hyde-Thompson}. However, there exist no orbifold theory for the classification of layer groups or for describing 2-periodic entanglement.

In this paper we introduce a new orbifold framework for the symmetry classification of three-dimensional layer groups.  While orbifold notation is well-established for the $17$ wallpaper groups of the Euclidean plane $\mathbb{R}^2$, no analogous symbolic description existed for the $80$ layer groups, whose actions occur on the thickened plane $\mathbb{R}^2\times \mathbb{R}$.  In Sec.~\ref{sec:wallpaper} we review the orbifold theory of wallpaper groups. In Sec.~\ref{sec:orbifolds_layer_groups} we extend the orbifold construction to the three-dimensional setting of the layer groups, identifying the irreducible symmetries. Mathematically, we obtain a compact and coordinate-independent notation that encodes the topology of the quotient space. 

After establishing this general layer orbifold framework, in Sec.~\ref{sec: fabrics} we illustrate its effectiveness through a detailed analysis of relevant knitted structures. Knitted fabrics provide a rich family of three-dimensional doubly periodic entangled structures exhibiting nontrivial layer-group symmetries.
Previous works on the intersection of knitted fabrics and mathematics have focused on knot theory~\cite{Shimamoto2025TopologicalDefectPropagation, MarkandeMatsumoto2020, Grishanov-knot-polynomials}. Here we focus on spatial symmetries of the three-dimensional shape of knitted fabric that emerge from coupling the underlying knot structure with the conditions necessary for mechanical equilibrium. The four basic knitted stitches are shown in Fig.~\ref{fig:four_stitches}, both their schematic structure and their presentation in fabric. To describe knitted fabrics beyond the simplest stitch, stockinette, we must consider the thin third dimension of the fabric. Here we consider these stitches through the lens of the extended orbifold framework developed in Sec.~\ref{sec:orbifolds_layer_groups} and find that this is an apt means to describe these materials' structures and symmetries.

\section{Orbifold of 2D Wallpaper Groups}\label{sec:wallpaper}

Doubly periodic tilings of the Euclidean plane $\mathbb{R}^2$ are classified by symmetry groups preserving the underlying pattern, namely their wallpaper symmetry groups. The isometries of $\mathbb{R}^2$ consist of translations, rotations, reflections, and glide reflections. In particular, a \textit{wallpaper group} is a discrete group of isometries of $\mathbb{R}^2$ containing two linearly independent translations. By the crystallographic restriction theorem, a wallpaper group may contain only rotations of order $2$, $3$, $4$, or $6$. It is known that there exist $17$ wallpaper groups. All classical results recalled in this section are detailed in~\cite{sym_of_things}.
 
Each wallpaper group admits a fundamental domain, and the action of the group identifies edges and vertices of this domain to produce a compact quotient space. A \textit{fundamental domain} is a minimal region of the pattern from which the entire tiling can be generated by applying the symmetries of the group. More intuitively, one may think of it as the smallest piece of the pattern, or smallest `tile', needed to cover the whole plane by applying the symmetries of the group without overlaps or gaps. Unlike a crystallographic unit cell, which is determined purely by translations, a fundamental domain also reflects rotational, reflectional, or glide symmetries.

Identifying the fundamental domain boundary produces a compact quotient space, in which rotation centers, mirror lines, and glide reflections appear as singular or non-orientable features. Following Thurston, an \textit{orbifold} is a generalization of a surface that records precisely this quotient structure encoding symmetry. For a wallpaper group $G$, the quotient $O = \mathbb{R}^2 / G$ is a compact two-dimensional orbifold. Classifying wallpaper groups is therefore equivalent to classifying the possible orbifold structures that arise from such quotients. 

The topology of the underlying surface $X_O$ of an orbifold $O$ is determined by its orientable or non-orientable genus together with the number of boundary components. By the classification of compact surfaces, its Euler characteristic is given by: 
\[
\chi(X_O) =
\begin{cases}
2 - 2g - b, & \text{if $X_O$ is orientable},\\[4pt]
2 - k - b, & \text{if $X_O$ is non-orientable},
\end{cases}
\tag{1}
\]
where $g$ is the number of orientable handles, $k$ the number of non-orientable (twisted) handles (crosscaps), and $b$ the number of boundary components. In the quotient of a wallpaper group, translations generate handles of $X_O$, mirrors produce the boundary components, and glide reflections introduce crosscaps.

In addition to the topology of the underlying surface $X_O$, an orbifold $O$ of a wallpaper group may carry singular points arising from rotational symmetries: cone points, corresponding to isolated rotational centers of order $n$, and corner reflectors, corresponding to rotational centers of order $n$ lying on mirror boundaries. These singularities modify the Euler characteristic of $X_O$: a cone point of order $n$ removes $(1 - \frac{1}{n})$ from $\chi(X_O)$, while a corner reflector of order $n$ removes $\frac{1}{2}(1 - \frac{1}{n})$ from $\chi(X_O)$.

Combining the topology of the underlying surface with the contributions of the singularities gives the Euler characteristic of the orbifold:
\[
\chi(O)
= \chi(X_O)
    - \sum_{i=1}^{r}\Bigl(1 - \frac{1}{a_i}\Bigr)
    - \frac{1}{2}\sum_{j=1}^{s}\Bigl(1 - \frac{1}{b_j}\Bigr),
\tag{2}
\]
where $a_1,\dots,a_r$ are the orders of the cone points and $b_1,\dots,b_s$ the orders of the corner reflectors.

Since $\mathbb{R}^2$ has Euler characteristic $0$, every wallpaper orbifold satisfies $\chi(O)=0$. In view of equations (1) and (2), the underlying surfaces $X_O$ arising as quotients of wallpaper groups are precisely the sphere $S^2$, the disk $D^2$, the real projective plane $\mathbb{P}_2= \mathbb{R}\mathrm{P}^2$, the torus $T^2$, the Klein bottle $K$, the annulus $\mathcal{A}=S^1 \times I$, and the Möbius band $Mb$.

Conway introduced a symbolic notation that encodes the topology and singular data of a two-dimensional orbifold. For wallpaper groups, the relevant symbols are: $*$ for the presence of mirror boundary, $n$ for a cone point or a corner reflector of order $n$ (where numbers appearing to the right of $*$ denote corner reflectors), $\times$ for a crosscap, and $o$ for the case with only translational symmetry.

In Conway’s combinatorial theory, each character in the orbifold symbol is assigned a numerical \textit{cost} $c$:
\[
\mathrm{cost}(o)=2,\qquad 
\mathrm{cost}(*)=1,\qquad 
\mathrm{cost}(\times)=1,
\]
\[
\mathrm{cost}(\text{cone }n)=1-\frac{1}{n},\qquad
\mathrm{cost}(\text{corner }n)=\frac{1}{2}\Bigl(1-\frac{1}{n}\Bigr),
\]
with cone and corner contributions matching the geometric terms appearing in equations~(1) and~(2). These symbolic costs are chosen so that the Euclidean condition $\chi(O)=0$ is simply encoded. The \textit{Magic Theorem} states that the symbols satisfying this constraint, that is, the symbols of total cost $2$, are precisely those that arise from wallpaper symmetry groups, and that each wallpaper group has a unique orbifold symbol.

\section{Orbifolds of 3D Layer Groups}\label{sec:orbifolds_layer_groups}

\subsection{Preliminaries on Layer Groups}

The layer groups are the three-dimensional analogues of the wallpaper groups: they describe the symmetries of patterns that are periodic in two directions (doubly periodic) but  not necessarily periodic in the transverse direction. They act by isometries in the thickened plane 
$\mathbb{R}^2 \times \mathbb{R}$, which we consider centered in the plane $z=0$. As in the planar case, a layer group admits a planar lattice of translations, but the additional transverse direction introduces symmetry operations not present in two dimensions. They form one of the crystallographic subperiodic families, consisting of three-dimensional crystallographic groups with only two independent translational directions. There are exactly $80$ layer groups~\cite{ITE-layer}.

\begin{figure}
\centering
\includegraphics[width=0.47\textwidth]{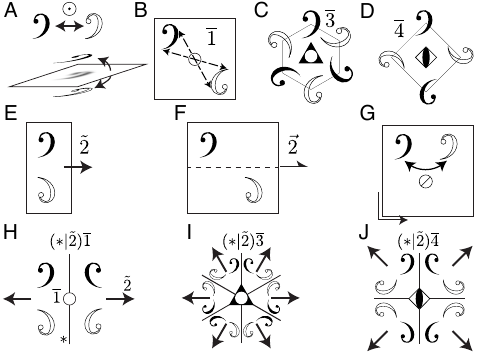}
\caption{\label{fig:symbols} (A) In-plane mirror: $\odot$; (B) Inversion center: $\overline{1}$; (C) 3-fold rotoinversion: $\overline{3}$; (D) 4-fold rotoinversion: $\overline{4}$; (E) In-plane 2-fold rotation: $\tilde{2}$;(F) In-plane 2-fold screw rotation: $\vec{2}$; (G) In-plane glide: $\oslash$; composite $(*|\tilde{2})$ of an in-plane mirror and in-plane 2-fold rotation at different angles, giving rise to an inversion (H), 3-fold rotoinversion (I), and 4-fold rotoinversion (J).}
\end{figure}

To encode these symmetry operations in a compact and topologically meaningful way, we introduce Conway-inspired symbols adapted to the layer-group setting. Each symbol encodes the corresponding orbifold feature in the quotient $(\mathbb{R}^{2}\times \mathbb{R})/L$, generalizing the role played by orbifold symbols in two dimensions. For vertical mirror planes, glide planes, rotations about axes perpendicular to the layer, and pure translations, we retain the planar symbols, namely $*$, $\times$, $n=2,3,4,6$, and $o$, respectively. Fig.~\ref{fig:symbols} illustrates the newly introduced symbols.  

These additional symbols arise from symmetry operations that have no analog in the planar wallpaper setting but become unavoidable in three dimensions. An in-plane $2$-fold axis may act either as a pure rotation with fixed horizontal axis, denoted by ($\tilde{2}$), or as a fixed-point-free screw rotation involving a half-translation along the axis,  denoted by $\vec{2}$. Inversion centers, denoted by $\bar{1}$, and roto-inversion centers, denoted by $\bar{n}= \bar{3}, \bar{4}, \bar{6}$, similarly appear as $z$-reversing point symmetries whose restriction to the plane $z=0$ act as a rotation, generating singularities similar to the planar cone points. Finally, mirrors and glides parallel to the layer must be marked separately from those perpendicular to it, because they contribute differently in the quotient orbifold. The symbol $\odot$ is used for in-plane mirror while the symbol $\oslash$ is used for in-plane glides.
We also introduce the composite symbol $(*|\tilde{2})$, denoting a vertical mirror plane meeting an in-plane 2-fold axis that is perpendicular to the mirror plane  at $z=0$. Their intersection is either an inversion point or a roto-inversion point.
Together, these symbols provide a complete description of all possible features of a layer-group action, extending Conway's notation.

Table~\ref{tab:layer-symbols-basic} lists the symbols used in this paper together with the corresponding isometries of $\mathbb{R}^2 \times  \mathbb{R}$. This notation forms the foundation for the orbifold description of layer groups developed in the subsequent sections. 

\begin{table}[h!]
\centering
\begin{tblr}{
  width=0.95\linewidth,
  colspec = {| l | X[l,m] |},
  colsep = 6pt,
  row{1} = {font=\small\bfseries},
  rowsep = 3pt
}
\hline
\textbf{Symbol} & \textbf{Corresponding isometry of $\mathbb{R}^2 \times \mathbb{R}$} \\
\hline
$o$ & No symmetry other than translations. \\
\hline
$\times$ & Vertical glide plane: 3D extension of a planar glide-reflection axis. \\
\hline
$*$ & Vertical mirror plane: 3D extension of a planar mirror axis. \\
\hline
$n=2,3,4,6$ & $n$-fold rotation about a cone axis or a dihedral edge perpendicular to the layer plane: 3D extension of planar cone points and corner reflectors.\\
\hline
$\tilde{2}$ & In-plane $2$-fold rotation: rotation about an axis parallel to the layer (acts as a reflection at $z=0$). \\
\hline
$\vec{2}$ & In-plane $2$-fold screw rotation: in-plane $2$-fold rotation combined with translation along the axis (acts as a glide reflection at $z=0$). \\
\hline
$\bar{1}$ &
Inversion center (acts as a 2-fold rotation at $z=0$). \\
\hline
$\bar{n} = \bar{3},\bar{4},\bar{6}$ &
Roto-inversion of order $n = 3,4,6$: $n$-fold rotation combined with inversion (acts as an $k$-fold rotation at $z=0$, with $k=6,4,3$ respectively).\\
\hline
$\odot$ &
In-plane mirror: mirror plane parallel to the layer. \\
\hline
$\oslash$ &
In-plane glide reflection: glide plane parallel to the layer. \\
\hline
\end{tblr}
\caption{Symbols and corresponding symmetry operations used for the orbifold notation of layer groups.}
\label{tab:layer-symbols-basic}
\end{table}

\subsection{Orbifolds of Layer Groups}

Let first $L$ be a crystallographic layer group acting on the thickened plane $\mathbb{R}^2 \times \mathbb{R}$, that is obtained as a direct extension of one of the 17 wallpaper group, namely one containing no symmetry that reverses the $z$-coordinate. The quotient $O = (\mathbb{R}^2 \times  \mathbb{R})/L$ is a three-dimensional orbifold. As in the planar setting, we denote by $X_O$ the underlying $3$–manifold obtained by forgetting the singular structure.

Thickening a wallpaper orbifold does not change the Euler characteristic of its underlying space. Since $ \mathbb{R}$ is contractible, we have $\chi( \mathbb{R})=1$, and therefore
\[
\chi(X_O)
=\chi(X_{O_W}\times  \mathbb{R})
=\chi(X_{O_W})\,\chi( \mathbb{R})
=\chi(X_{O_W}),
\]
where $O_W=\mathbb{R}^2/W$ is the orbifold of a wallpaper group $W$ with underlying surface $X_{O_W}$. Thus for such layer groups, the underlying manifold of a layer orbifold has the same Euler characteristic as the underlying surface of its wallpaper planar projection, given by equation (1) of the previous section. In particular, the underlying $3$–manifold $X_O$ is one of the thickened surfaces arising from the planar wallpaper projection, namely: the thickened sphere: $S^2 \times \mathbb{R}$, the thickened disk: $D^2 \times \mathbb{R}$, the thickened projective plane: $\mathbb{P}_2 \times \mathbb{R}$, the thickened torus: $T^2 \times \mathbb{R}$, the thickened Klein bottle: $K \times \mathbb{R}$, the thickened annulus: $ \mathcal{A} \times \mathbb{R}$, and the thickened Möbius band: $\mathrm{Mb}\times \mathbb{R}$. The plain knit of Sec.~\ref{sec: fabrics} is an example. 

Consider now layer groups that are not direct extensions of wallpaper groups. In these cases, the layer group contains at least one element that reverses the $z$-coordinate, such as an in-plane reflection or glide reflection, an in-plane 2-fold rotation or screw rotation, an inversion or a roto-inversion symmetry. The presence of such $z$-reversing symmetries affects the structure of the quotient: the symmetry action remains well defined in $\mathbb{R}^2 \times \mathbb{R}$ but the resulting orbifold is, in general, no longer a simple thickening of a planar orbifold. 

The 17 layer groups that contain in-plane reflections are quotients of the above 17 wallpaper-type groups by a mirror symmetry at $z=0$, thus creating a boundary. Since this boundary plane divides the layer $(R^2 \times R)$ into two identical half-spaces, the underlying topological space $X_O$ of the associated orbifold becomes $X_{O_W} \times [0,\infty)$, using the same notation as above.

For each of the remaining 46 layer groups, the study of the underlying topological space $X_O$ appears to be more complex. 
The examples of garter, rib and seed stitch discussed in Sec.~\ref{sec: fabrics} illustrate this complexity. The number of boundary components arising from mirror planes in $X_O$ can be deduced from the set of symmetries of the corresponding layer group. The orientability of $X_O$ is influenced by the presence of non-mirror orientation-reversing identifications. Glide reflections, inversions, and roto-inversions, whose linear parts have determinant $-1$, usually lead to a non-orientable underlying space, while pure mirrors do not reverse the orientability of $X_O$. The orbifold symbols presented in the fourth column of Tables~\ref{tab:layer-symbols-table1} and~\ref{tab:layer-symbols-table2} encode the boundary components and orientability, as described in the next section. 

The additional properties induced by the singular locus are more difficult to capture in a simple way. Since these 46 layer groups contain $z$-reversing symmetries, points with opposite transverse coordinates are identified through the group action. The quotient space can thus be represented using the half-thickness $\mathbb{R}^2 \times [0, \infty)$, with appropriate identifications that may differ for $z>0$ and for $z=0$. For $z\neq0$, a $z$-reversing symmetry does not fix any point, so the local stabilizers are determined only by the $z$-preserving symmetries, such as vertical reflections and rotations. However, the plane $z=0$  may be further quotiented by the action of the $z$-reversing symmetries, giving rise to additional singular strata in the orbifold. 
The underlying spaces of these layer-group orbifolds are thus, in general, not uniform product spaces and should be seen as stratified quotient spaces. 
A precise description of these complex spaces is non-trivial and will be investigated further in future work.

\subsection{The Orbifold Symbols of Layer Groups}

We now describe the $80$ layer groups using our new orbifold notation. This provides a symbolic system that complements the traditional Hermann–Mauguin (HM) notation. The HM notation, introduced to standardize crystallographic symmetry descriptions, encodes the geometric symmetry operations acting on a chosen conventional unit cell. While this framework is important for crystallography, it does not record the topology of the quotient space obtained from the group action. Our orbifold notation is an attempt to fill this gap by encoding the topology and singularities of a fundamental domain of a layer group.
This framework is well suited for modeling and classifying 3D doubly periodic structures, where the topology of the underlying network is a key descriptor alongside the crystallographic symmetry.

The symmetry operations and properties of the layer groups are detailed in the \textit{International Tables for Crystallography, Volume E} (Kopský \& Litvin, 2010), hereafter referred to as ITE~\cite{ITE-layer}. Each layer group is identified by a unique number and by its HM symbol, which form the first two columns of our classification  Tables~\ref{tab:layer-symbols-table1} and~\ref{tab:layer-symbols-table2}. The initial letter of the HM symbol specifies the centering type of the conventional unit cell (primitive $p$ or centered $c$), while the following characters describe the symmetry elements acting along the principal directions.

Every layer group admits a canonical planar projection obtained by forgetting the transverse structure. This projection preserves all symmetry features that act within the layer, while collapsing purely three-dimensional features such as in-plane mirrors, in-plane glides, in-plane rotations, or screw rotations parallel to the layer into their two-dimensional counterparts. This projection is used solely to relate the layer group to its 2D planar orbifold, whose Euler characteristic and symbolic cost were analyzed in the preceding subsections. The third column of  Tables~\ref{tab:layer-symbols-table1} and~\ref{tab:layer-symbols-table2} record the 2D wallpaper group corresponding to the  projection of the layer group, given both in HM notation and in the associated Conway symbol of the resulting two-dimensional orbifold. 

The fourth column of Tables~\ref{tab:layer-symbols-table1} and~\ref{tab:layer-symbols-table2} list our new orbifold symbol for each layer group. This symbol is determined  by the orbifold features of the action of the group on the thickened plane, independently of any coordinate system, and each layer group has a different symbol. Note that the polygon corresponding to the flat orbifold of a layer group may sometimes differ from the asymmetric unit as given in the ITE. 

For each layer group $L$ with its corresponding 3-orbifold $O=(\mathbb{R}^2 \times \mathbb{R}) / L$, a canonical orbifold symbol is assigned to encode the singular strata of $O$, as well as any additional fixed-point-free features needed to determine the topology of $O$. To capture the quasi-two-dimensionality of structures exhibiting layer group symmetries, these 3-orbifold symbols are constructed by extending Conway's rules for 2-orbifold symbols to the three-dimensional setting, replacing rotation centers by axes and (glide) reflection lines by planes. These rules are also adapted to the additional symbols introduced here for ordering purposes: $\bar{1}, \bar{3}, \bar{4}, \bar{6}$ are treated like the numbers $2,3,4,6$ since the induced two-dimensional action of the corresponding symmetries acts like rotation centers, while the symbol $\tilde{2}$ is treated like the symbol $*$ since the induced two-dimensional action of the corresponding symmetry acts like a mirror line, as noted in Table~\ref{tab:layer-symbols-basic}. Unlike Conway’s 2-orbifold symbols for wallpaper groups, these 3-orbifold symbols may not encode a description of all the Wyckoff sites as listed in the ITE~\cite{ITE-layer}. To keep the notation minimal, any symmetry that arises as the composition of other symmetries encoded in the symbol is omitted. The choice of symbols is described further below. We summarize the layer group orbifold symbol construction rules as follows:
\begin{enumerate}
    \item If $L$ contains an in-plane reflection symmetry, then the corresponding symbol $\odot$, which indicates the presence of a mirror plane boundary, should appear in the first position and cannot be permuted with any other symbol.
    
    \item If $L$ contains an in-plane glide reflection symmetry which cannot be recovered by the composition of symmetries already encoded in the orbifold symbol, then the corresponding symbol $\oslash$ should appear in the first position and cannot be permuted with any other symbol. Note that the groups containing the symbol $\oslash$ do not contain the symbol $\odot$.
    
    \item The symbols $2,3,4,6$, as well as the symbols $\bar{1}, \bar{3}, \bar{4}, \bar{6}$, that are not preceded by a symbol $*$, $\tilde{2}$ or $(*|\tilde{2})$ can be freely permuted.

    \item The symbols $2,3,4,6$, as well as the symbols $\bar{1}, \bar{3}, \bar{4}, \bar{6}$, directly following a symbol $*$, $\tilde{2}$ or $(*|\tilde{2})$ can be cyclically permuted.
    
    \item A block of type $*A \cdots B$ can be freely permuted with a block of type $\tilde{2} \, C \cdots D$ where the letters represent rotations, inversions, or roto-inversions: $*A \cdots B \ \tilde{2} \, C \cdots D = \tilde{2} \, C \cdots D \ *A \cdots B$.

    \item A block of type $(*|\tilde{2}) \, A \cdots B$ should appear after a block of type $*C \cdots D$ or $\tilde{2} E \cdots F$, if applicable to $O$, and cannot be permuted with them. Namely, we can only have: $*C \cdots D \ \tilde{2} \, E \cdots F  \ (*|\tilde{2}) \, A \cdots B = \tilde{2} \, E \cdots F \ *C \cdots D$ $(*|\tilde{2}) \, A \cdots B$. In this case, the symbols $*$ and $\tilde{2}$ in $(*|\tilde{2}) \, A \cdots B$ represent singular features already encoded by the $*C \cdots D$ or $\tilde{2} E \cdots F$ blocks, if applicable to $O$, and thus do not add any additional singularity.

    \item If $O$ can be assigned different orbifold symbols satisfying the above rules, then the choice can be made unique by imposing the following ordering based on the dimensionality: boundary components are prioritized over other singularities, and these are prioritized over symmetries that do not fix points.
\end{enumerate}

This proposed orbifold notation should be seen as a descriptive tool reflecting the stratified structure of the quotient space, rather than as a complete invariant determining the layer group uniquely. A concrete example is presented in Section~\ref{sec: fabrics} (see garter stitch in Fig. 5).

\begin{table}[h!]
\centering
\begin{tblr}{
  width = \linewidth,
  colspec = {c c c c},
  colsep = 3pt,
  row{1} = {font=\small\bfseries},
  rowsep = 2pt,
  column{1} = {wd=5mm},
  column{2} = {wd=20mm},
  column{3} = {wd=25mm},
  column{4} = {wd=25mm},
  hlines,
  vlines,
}
N & HM symb. & 2D W.P. & Orb. symb. \\
1 
& $p1$ 
& $ p1 \ (o)$ 
& $o$ 
\\
2 
& $p\overline{1}$ 
& $ p2 \ (2222)$ 
& $\bar{1} \bar{1} \bar{1} \bar{1}$ 
\\
3 
& $p112$ 
& $p2 \ (2222)$ 
& $2222$ 
\\
4 
& $p11m$ 
& $ p1 \ (o)$ 
& $\odot$ 
\\
5 
& $p11a$ 
& $ p1 \ (o)$ 
& $\oslash$
\\
6 
& $p112/m$ 
& $p2 \ (2222)$ 
& $\odot 2222$ 
\\
7 
& $p112/a$ 
& $p2 \ (2222)$ 
& $\bar{1} \bar{1}22$ 
\\
8 
& $p211$ 
& $ pm \ (**)$ 
& $\tilde{2} \tilde{2}$ 
\\
9 
& $p2_111$ 
& $pg \ (\times\times)$ 
& $\vec{2} \vec{2} $ 
\\
10 
& $c211$ 
& $cm \ (*\times)$ 
& $\tilde{2} \vec{2}$ 
\\
11 
& $pm11$ 
& $pm \ (**)$ 
& $**$ 
\\
12 
& $pb11$ 
& $pg \ (\times\times)$ 
& $\times\times$ 
\\
13 
& $cm11$ 
& $cm \ (*\times)$ 
& $*\times$ 
\\
14 
& $p2/m11$ 
& $pmm \ (*2222)$ 
& $(*|\tilde{2})\bar{1}\bar{1}\bar{1}\bar{1}$
\\
15 
& $p2_1/m11$ 
& $pmg \ (22*)$ 
& $\bar{1} \bar{1}*$ 
\\
16 
& $p2/b11$ 
& $pmg \ (22*)$ 
& $\bar{1} \bar{1} \tilde{2}$ 
\\
17 
& $p2_1/b11$ 
& $pgg \ (22\times)$ 
& $\bar{1} \bar{1} \times$ 
\\
18 
& $c2/m11$ 
& $cmm \ (2*22)$ 
& $\bar{1}(*|\tilde{2})\bar{1}\bar{1}$
\\
19 
& $p222$ 
& $pmm \ (*2222)$ 
& $\tilde{2}2222$ 
\\
20 
& $p2_122$ 
& $pmg \ (22*)$ 
& $22\tilde{2}$ 
\\
21 
& $p2_12_12$ 
& $ pgg \ (22\times)$ 
& $22\vec{2}$ 
\\
22 
& $c222$ 
& $cmm \ (2*22)$ 
& $2 \tilde{2} 22$ 
\\
23 
& $pmm2$ 
& $pmm \ (*2222)$ 
& $*2222$ 
\\
24 
& $pma2$ 
& $pmg \ (22*)$ 
& $22*$ 
\\
25 
& $pba2$ 
& $pgg \ (22\times)$ 
& $22\times$ 
\\
26 
& $cmm2$ 
& $cmm \ (2*22)$ 
& $2*22$ 
\\
27 
& $pm2m$ 
& $pm \ (**)$ 
& $\odot**$
\\
28 
& $pm2_1b$ 
& $pm \ (**)$ 
& $\oslash**$ 
\\
29 
& $pb2_1m$ 
& $pg \ (\times\times)$ 
& $\odot\times\times$ 
\\
30 
& $pb2b$ 
& $pm \ (**)$ 
& $\oslash \tilde{2} \tilde{2}$
\\
31 
& $pm2a$ 
& $pm \ (**)$ 
& $* \tilde{2}$
\\
32 
& $pm2_1n$ 
& $cm \ (*\times)$ 
& $* \vec{2}$
\\
33 
& $pb2_1a$ 
& $pg \ (\times\times)$ 
& $\times \vec{2}$
\\
34 
& $pb2n$ 
& $cm \ (*\times)$ 
& $\tilde{2} \times$
\\
35 
& $cm2m$ 
& $cm \ (*\times)$ 
& $\odot * \times$ 
\\
36 
& $cm2e$ 
& $pm \ (**)$ 
& $\oslash * \tilde{2}$
\\
37 
& $pmmm$ 
& $pmm \ (*2222)$ 
& $\odot*2222$ 
\\
38 
& $pmaa$ 
& $pmm \ (*2222)$ 
& $\tilde{2} 22 (*|\tilde{2}) \bar{1}\bar{1}$
\\
39 
& $pban$ 
& $cmm \ (2*22)$ 
& $\bar{1}\tilde{2} 22$
\\
40 
& $pmam$ 
& $pmg \ (22*)$ 
& $\odot 22*$ 
\\
\end{tblr}
\caption{Orbifold classification table for layer groups (1-40) 
listing: the ITE number (N), its HM symbol, the wallpaper projection (2D W.P.), and the orbifold symbol.
}
\label{tab:layer-symbols-table1}
\end{table}

\begin{table}[h!]
\centering
\begin{tblr}{
  width = \linewidth,
  colspec = {c c c c},
  colsep = 3pt,
  row{1} = {font=\small\bfseries},
  rowsep = 2pt,
  column{1} = {wd=5mm},
  column{2} = {wd=20mm},
  column{3} = {wd=25mm},
  column{4} = {wd=25mm},
  hlines,
  vlines,
}
N & HM symb. & 2D W.P. & Orb. symb.\\
41 
& $pmma$ 
& $pmm \ (*2222)$ 
& $*22 (*|\tilde{2}) \bar{1}\bar{1}$
\\
42 
& $pman$ 
& $cmm \ (2*22)$ 
& $2 (*|\tilde{2}) \bar{1}\bar{1}$
\\
43 
& $pbaa$ 
& $pmg \ (22*)$ 
& $2 \bar{1} \tilde{2}$
\\
44 
& $pbam$ 
& $pgg \ (22\times)$ 
& $\odot22\times$ 
\\
45 
& $pbma$ 
& $pmg \ (22*)$ 
& $2 \bar{1} *$
\\
46 
& $pmmn$ 
& $cmm \ (2*22)$ 
& $\bar{1} * 22$
\\
47 
& $cmmm$ 
& $cmm \ (2*22)$ 
& $\odot2*22$ 
\\
48 
& $cmme$ 
& $pmm \ (*2222)$ 
& $*2 \tilde{2}2 (*|\tilde{2}) \bar{1}\bar{1}$
\\
49 
& $p4$ 
& $p4 \ (442)$ 
& $442$ 
\\
50 
& $p\bar{4}$ 
& $p4 \ (442)$ 
& $\bar{4}\bar{4}2$ 
\\
51 
& $p4/m$ 
& $p4 \ (442)$ 
& $\odot442$ 
\\
52 
& $p4/n$ 
& $p4 \ (442)$
& $4 \bar{4} \bar{1}$
\\
53 
& $p422$ 
& $p4m \ (*442)$ 
& $\tilde{2}442$ 
\\
54 
& $p42_12$ 
& $ p4g \ (4*2)$ 
& $4\tilde{2}2$ 
\\
55 
& $p4mm$ 
& $p4m \ (*442)$ 
& $*442$ 
\\
56 
& $p4bm$ 
& $ p4g \ (4*2)$ 
& $4*2$ 
\\
57 
& $p\bar{4}2m$ 
& $ p4m \ (*442)$ 
& $\tilde{2} 2 (*|\tilde{2}) \bar{4} \bar{4}$
\\
58 
& $p\bar{4}2_1m$ 
& $p4g \ (4*2)$ 
& $\bar{4}*2$ 
\\
59 
& $p\bar{4}m2$ 
& $p4m \ (*442)$ 
& $* 2 (*|\tilde{2}) \bar{4} \bar{4}$
\\
60 
& $p\bar{4}b2$ 
& $p4g \ (4*2)$ 
& $\bar{4} \tilde{2} 2$
\\
61 
& $p4/mmm$ 
& $ p4m \ (*442)$ 
& $\odot*442$ 
\\
62 
& $p4/nbm$ 
& $p4m \ (*442)$ 
& $\tilde{2} 4 (*|\tilde{2}) \bar{4} \bar{1}$
\\
63 
& $p4/mbm$ 
& $p4g \ (4*2)$ 
& $\odot4*2$ 
\\
64
& $p4/nmm$ 
& $p4m \ (*442)$ 
& $* 4 (*|\tilde{2}) \bar{4} \bar{1}$
\\
65 
& $p3$ 
& $p3 \ (333)$ 
& $333$ 
\\
66 
& $p\bar{3}$ 
& $p6 \ (632) $ 
& $\bar{3}3\bar{1}$ 
\\
67 
& $p312$ 
& $p3m1 \ (*333)$ 
& $\tilde{2}333$ 
\\
68 
& $p321$ 
& $p31m \ (3*3)$ 
& $3\tilde{2}3$ 
\\
69 
& $p3m1$ 
& $p3m1 \ (*333)$ 
& $*333$ 
\\
70 
& $p31m$ 
& $p31m \ (3*3)$ 
& $3*3$ 
\\
71
& $p\bar{3}1m$ 
& $p6m \ (*632)$ 
& $ \tilde{2} 3 (*|\tilde{2}) \bar{3} \bar{1}$
\\
72 
& $p\bar{3}m1$ 
& $p6m \ (*632)$ 
& $* 3 (*|\tilde{2}) \bar{3} \bar{1}$
\\
73 
& $p6$ 
& $p6 \ (632)$ 
& $632$ 
\\
74 
& $p\bar{6}$ 
& $p3 \ (333)$ 
& $\odot 333$
\\
75 
& $p6/m$ 
& $p6 \ (632)$ 
& $\odot632$ 
\\
76 
& $p622$ 
& $p6m \ (*632) $ 
& $\tilde{2}632$ 
\\
77 
& $p6mm$ 
& $p6m \ (*632)$ 
& $*632$ 
\\
78 
& $p\bar{6}m2$ 
& $p3m1 \ (*333)$ 
& $\odot *333$ 
\\
79 
& $p\bar{6}2m$ 
& $p31m \ (3*3)$ 
& $\odot 3*3$ 
\\
80 
& $p6/mmm$ 
& $p6m \ (*632)$ 
& $\odot *632$ 
\\
\end{tblr}
\caption{Orbifold classification table for layer groups (41-80) 
listing: the ITE number (N), its HM symbol, the wallpaper projection (2D W.P.), and the orbifold symbol.
}
\label{tab:layer-symbols-table2}
\end{table}

\section{Symmetries of Knitted Fabrics}\label{sec: fabrics}

We now consider the symmetries in a selection of common textile structures: the four basic weft-knitted stitches. Just as visual arts such as mosaics, tapestries, and printed patterns illustrate wallpaper groups on their respective canvases ~\cite{Grunbaum2006}, knitted and woven textiles depict layer symmetries. However, the full richness of these symmetries are hidden within the thin transverse thickness of the material, with hints of the underlying structure on display as a fabric texture, which gives the impression of wallpaper symmetries.

\begin{figure}[h]
    \centering  \includegraphics[width=\linewidth]{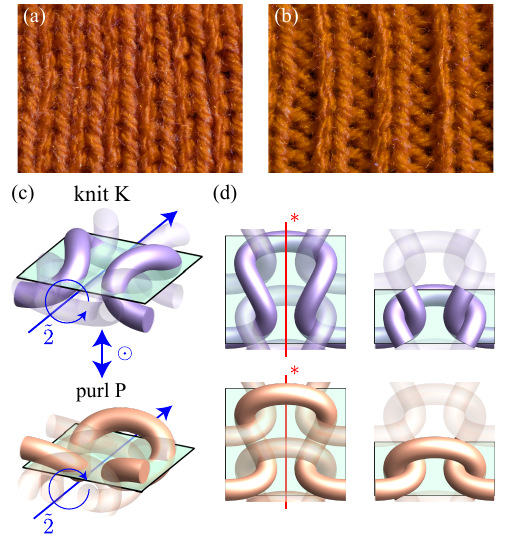}
    \caption{(a) Rib fabric with no applied forces resembles stockinette fabrics. (b) When rib fabric is stretched in the horizontal direction, its three dimensional structure and layer symmetries are revealed. (c) Knit and purl stitches are related by an in-plane mirror \textit{or} a 2-fold rotation about the axis parallel to the wale direction. (d) Top-down projections of knit and purl stitches; in the case of translational periodicity, each loop can be reduced, given a set of periodic boundaries in the shaded region (right).}
    \label{fig:stitch_overview}
\end{figure}

Perhaps one of the clearest consequences of layer symmetries is through the distinctive anisotropic mechanical response of rib-knit fabric. The stripe-based texture of rib fabric appears similar to plain-knit (stockinette or jersey) at a first glance. If the material is stretched course-wise then the front stripes separate, showing a striped reversed pattern that alternates with the front-facing pattern. These two strikingly different states of the same fabric are shown in Fig.~\ref{fig:stitch_overview}a, b.
Notably, this striped patterning gives rise to a soft, highly-extensile response when stretched along the course but a much stiffer response along the wale~\cite{Singal2023}.

While there is a variety of stitches that can be employed by a knitter, the most basic are the `knit' (K) and the `purl' (P).
The knit stitch, shown in fig.~\ref{fig:stitch_overview}c, joins neighboring courses (rows) of yarn through a slip knot that loops behind the fabric, whereas the purl has a loop that lies in front of the fabric.
A separate reduced representation, shown in the right panels of fig.~\ref{fig:stitch_overview}d, defines a stitch as the pair of claps formed by segments of two separate courses of yarn. As we explore below, this reduced representation is equivalent to the former when stitch symmetry permits the use of periodic stitch boundaries to identify free ends of each yarn segment.

Each stitch, if excised from the surrounding fabric, only contains a single point symmetry: a mirror plane that cuts transverse to the plane of the fabric and lies along the wale direction ($*$ in the orbifold notation).
Any `rectangular' swatch of purely K/P-patterned fabric whose stitches can be organized by a direct product of $m$ wales and $n$ courses maintains these mirror planes along the wales.
Furthermore, the K and P stitches are related by either a mirror operation through the plane of the fabric (i.e., an action of $\odot$) or through a $\pi$-rotation of the fabric about the wale axis (i.e., an action of $\tilde{2}$), as shown in Fig.~\ref{fig:stitch_overview}c,d.
In the case of circular knitting, where the swatch forms a cylinder, allowing for unambiguous definition of a front and reverse side of the fabric, K and P are related by a mirror operation through the fabric.
However, for flat knitting, the yarn direction is reversed between subsequent courses. To implement this reversal in hand knitting, the needles are exchanged between the hands. Thus, in this case K and P are related by a rotation $\pi$ around the wale axis.
Note that the two operations that interconvert between K and P are identical only if the stitches maintain their wale-wise mirrors.
When this symmetry is broken, as happens when the yarn has a preferred twist~\cite{PavkoCuden2015,Inui2017}, care must be taken in how to relate K and P.
Note, however, the incorporation of yarn twist requires the incorporation of curve framing (i.e.~through prescription of a material frame). This will modify the symmetries present in the stitch thus we leave consideration of yarn twist to future studies.
Other ways to remove mirror symmetries involve twisted variations of these stitches or altering the swatch lattice so that it is no longer a direct product of wale and course directions.
This latter change can be accomplished, for example, by knitting two stitches in one row together to form a single stitch in the next row, or by using short rows, resulting in defect structures analogous to edge dislocations in crystals.

\begin{figure*}
\centering
\includegraphics[width=\textwidth]{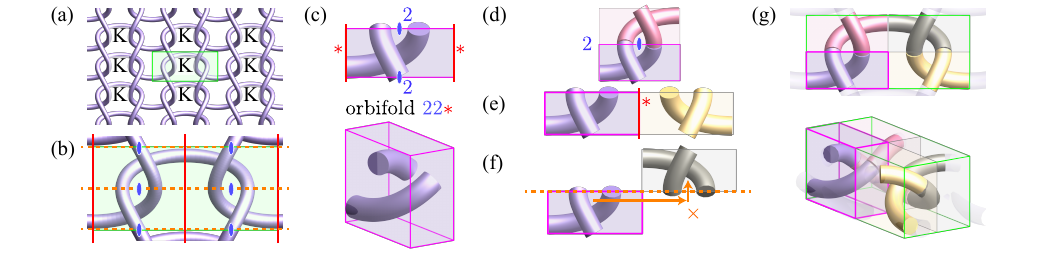}
\caption{\label{fig:stockinette} (a) Representation of the plain knit as a doubly-periodic tiling of space using just knit (K) stitches. The unit cell is boxed in green.(b) Close-up of the translational unit cell, highlighting two-fold axes (blue ellipses), perpendicular mirrors (red solid lines), and perpendicular glide planes (orange dashed lines). (c) Fundamental domain, showing boundary component $*$ of the orbifold. Illustrations of actions on the fundamental domain: (d) a two-fold rotation, (e) perpendicular mirror, and (f) perpendicular glide. (g) Translational unit cell (outlined in green) tiled by fundamental domains (outlined in pink). }
\end{figure*}

In many ways, the simplest knitted fabric is the plain knit, also known as the stockinette or jersey, shown in the upper left quadrant of Fig. \ref{fig:four_stitches} and Fig.~\ref{fig:stockinette}a. When viewed from the front of the fabric, in the knitting convention, a plain knit is a rectangular lattice of K stitches (fig.~\ref{fig:stockinette}a).
We will limit our discussion to infinite patterns, considering fabric regions far from the boundaries, known as cast on, bind off, edges, and joins in knitting parlance.
In this case, the wale-wise mirror (symbol: $*$) is joined by a pair of new two-fold axes (symbol: 2) that pierces through the plane of the fabric.
These separate two-fold axes are centered at the center of a yarn-yarn crossing or `clasp' as well as the midpoints of the wale-wise connecting segments of yarn.
The mirror symmetry, combined with each of these two-fold axes, creates a pair of glide planes through the fabric. These symmetry operations are shown in Fig.~\ref{fig:stockinette}b. As a result, the layer group is identified as pma2, in agreement with~\cite{OKeeffeTreacy2022Isogonal}. The fundamental domain of the fabric is identified as half of a yarn-yarn clasp, shown in Fig.~\ref{fig:stockinette}c. Focusing on the mirror along with the pair of two-fold axes as fundamental generators of the symmetry group, we find that the orbifold signature is $22*$, which is consistent with the layer group identification.
As shown in Fig.~\ref{fig:stockinette}d-g, the full translational unit cell can be constructed from the symmetry elements acting on the fundamental domain.
Note that both the layer group and orbifold identification mark the plane knit as a structure that has purely 2D symmetries---that is, pma2 is also a wallpaper group.

\begin{figure}[h!]
\centering
\includegraphics[width=\linewidth]{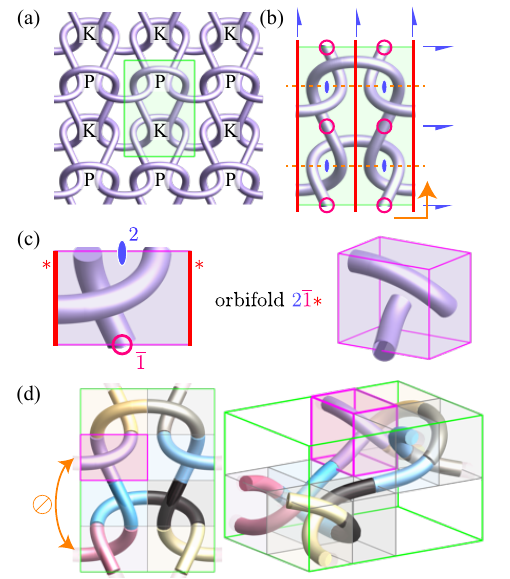}
\caption{\label{fig:garter}  (a) Representation of the garter stitch as a doubly-periodic tiling of space using alternating courses of knit (K) and purl (P) stitches. (b) Close-up of the translational unit cell, highlighting two-fold axes (blue ellipses), perpendicular mirrors (red solid lines), perpendicular glide planes (orange dashed lines), inversion centers (pink circles), the parallel glide symmetry (orange arrow), and in-plane screw axes (blue half-arrows). (c) Orbifold showing boundary component $*$ and symmetry elements $2$ and $\overline{1}$. (d) Translational unit cell tiled by fundamental domains, along with a depiction of an in-plane glide.}
\end{figure}

In garter stitch, a course of K alternates with a course of P, as shown in the upper right quadrant of Fig.~\ref{fig:four_stitches} and Fig.~\ref{fig:garter}a. We analyze the case where the fabric is slightly stretched in the wale direction, revealing the full texture of the fabric in a manner analogous to the unfolding of the rib fabric shown in Fig.~\ref{fig:stitch_overview}a,b. In this rectangular fabric setting, we can regard that P stitch as being obtained from K by translation along the wale by a single stitch, half of a translational unit, followed by an in-plane reflection.
As a result, the fabric has a glide symmetry parallel to the plane, which we denote by $\oslash$.
Similar to the plain knit, the garter has two-fold axes in the centers of the yarn-yarn clasp regions.
However, as shown in Fig.~\ref{fig:garter}b, the new parallel glide symmetry converts the two-fold axis placed midway along the wale-wise connecting yarn segments into inversion centers, denoted $\overline{1}$.
As a result there is only a single independent glide plane passing perpendicular through the fabric along the course direction.
The combination of wale-wise mirror and inversion center generates parallel screw axes (denoted $\vec{2}$) along the wale direction, with axes collinear with the mirror planes, along with separate parallel screws along the course direction, aligned with the perpendicular glide planes. 
We identify the orbifold (Fig.~\ref{fig:garter}c), which is bounded by pairs of mirrors and two-fold axes.
As shown in Fig.~\ref{fig:garter}e, this orbifold can generate the remainder of the translational unit cell.
However, we emphasize that the orbifold is not the `minimal unit' of the structure: a smaller asymmetric unit can be created under a further quotient of the orbifold by the inversion operation $\overline{1}$, as shown in Fig.~\ref{fig:garter}d.
Note that the inclusion of this operation is incompatible with the definition of an orbifold as it would only consider the upper-half of the fundamental domain. 
We identify garter as orbifold $2\overline{1}*$, and hence layer group pbma (in agreement with~\cite{OKeeffeTreacy2022Isogonal}). 

\begin{figure}[h!]
\centering
\includegraphics[width=.5\textwidth]{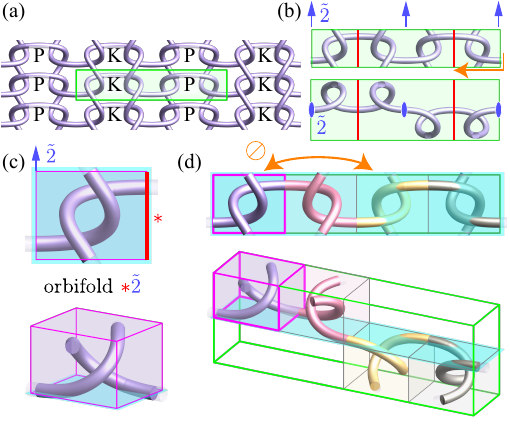}
\caption{\label{fig:rib}  (a) Representation of the rib stitch as a doubly-periodic tiling of space using alternating wales of knit (K) and purl (P) stitches. (b) Close-up of the translational unit cell, the above schematic in-plane and below schematic with horizontal axis along the course and vertical axis along the thickness of the fabric, highlighting perpendicular mirrors (red solid lines), the parallel glide symmetry (orange arrow), and in-plane two-fold axes (blue arrows/ellipses). (c) Fundamental domain showing orbifold boundary component $*$ and singularity $\tilde{2}$. (d) Translational unit cell tiled by fundamental domains, along with a depiction of an in-plane glide.}
\end{figure}

\begin{figure}[h!]
\centering
\includegraphics[width=\linewidth]{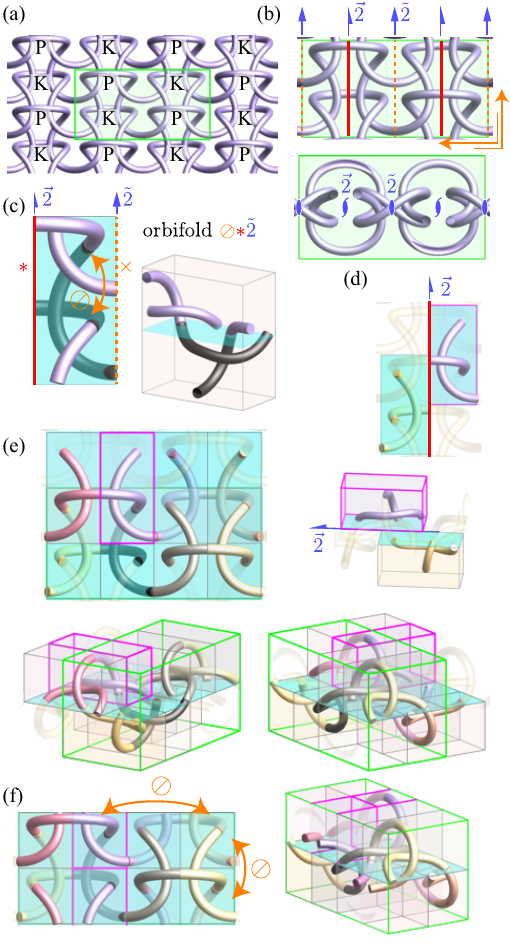}
\caption{\label{fig:seed} (a) Representation of the seed stitch using a checkerboard pattern of knit (K) and purl (P) stitches. (b) Top-down and side views of the translational unit cell, highlighting perpendicular mirrors (red solid lines), perpendicular glide planes (orange arrows), the parallel glide symmetries (orange dashed lines), in-plane two-fold axes (blue arrows/ellipses), and in-plane screw axes (blue half-arrows/swirls). (c) Illustration of orbifold with boundary component $*$ and singularity $\tilde{2}$, along with in-plane glide $\oslash$. (d) Illustration of the in-plane screw acting on a fundamental domain. (e) Construction of a translational unit cell tiled by fundamental domains; note that half of the domains involve a shift through the unit cell boundaries due to the action of the in-plane screw. (f) Wrapping of the fundamental domains into the conventional unit cell.}
\end{figure}

Next, consider our initial example of the stretched rib pattern, which is a relative of garter where wales alternate between pure K and P, shown in the lower right quadrant of Fig.~\ref{fig:four_stitches} and Fig.~\ref{fig:rib}a.
While similar to garter, rib is characterized only by perpendicular mirrors, an in-plane glide, and in-plane two-fold rotations (see Fig.~\ref{fig:rib}b).
This is because the rib pattern lacks the perpendicular two-fold axes of plain knit and garter. The fundamental domain, shown in Fig.~\ref{fig:rib}c, therefore extends the full width of the translational unit cell along the wale direction, adopting a single set of periodic boundaries.
Finally, much like the garter pattern, the resulting rib structure adopts distinct top and bottom layer components, as shown in Fig.~\ref{fig:rib}d. 
We identify rib as the orbifold $*\tilde{2}$, corresponding to layer group pm2a. It is interesting to note that the this presentation of the rib knit as a stacked structure with only half of each layer filled relies on the course-stretched configuration shown here; the relaxed rib structure in Fig.~\ref{fig:stitch_overview}a fills in these voids.

Our final example is seed, shown in the lower left quadrant of Fig.~\ref{fig:four_stitches}, which consists of an alternating `checkboard' pattern of K and P (see Fig.~\ref{fig:seed}a). This has the largest unit cell of the typical knit patterns, consisting of a pair of Ks and pair of Ps. As depicted in Fig.~\ref{fig:seed}b, the checkerboard structure gives rise to orthogonal pairs of in-plane glide symmetries. 
The remaining symmetries are all wale-oriented, much like rib, and consists of in-plane screw axes aligned with the perpendicular mirrors, along with in-plane two-fold axes aligned with perpendicular glide planes. The fundamental domain for seed is perhaps the least intuitive, consisting of a single layer of curve segments that belong to separate knit and purl stitches. 
Much like the rib stitch, the seed orbifold has a boundary component $*$ and a singularity $\tilde{2}$. 
The key difference lies in the in-plane glide reflection $\oslash$ perpendicular to the boundary component. We see that applying this in-plane glide to the purple yarn segments in Fig.~\ref{fig:seed}c connects these previously disjoint yarns with the black yarn segments in Fig.~\ref{fig:seed}c, forming the orbifold for seed.
However, since the screw symmetry involves a half-translation along the wale direction (Fig.~\ref{fig:seed}d) the image of this domain not only lies in the complementary layer, but is similarly translated by half a unit cell along the wale direction. As a result the image of the highlighted fundamental domain under symmetry actions gives rise to a distinctly staggered translational unit cell (Fig.~\ref{fig:seed}e). In accordance with the convention of constructing convex polyhedral translational units, the portions of fundamental domains that hang outside of the conventional rectangular (i.e.~extruded rectangular cuboid) unit cell are wrapped through the unit cell boundary, giving rise to the structure in Fig.~\ref{fig:seed}f. 
Since the in-plane screw, in-plane glide, and mirror operations generate the translational unit, we identify the orbifold as $\oslash*\tilde{2}$, which is layer group cm2e (in agreement with~\cite{takano2025}).

Each of the example structures shown in Figs.~\ref{fig:stockinette}-\ref{fig:seed} are `ideal', obtained by numerically minimizing a functional consisting of an elastica-style bending cost and a soft, repulsive inter-yarn interaction energy at a fixed length of yarn per stitch~\cite{Singal2023}.
Additionally, since the rib pattern contracts in the course-wise direction in the absence of external forces, a slight course-wise uniaxial stress was applied in order to reduce the overlap between knit and purl segments, making it easier to visualize the rib structure, seen in Fig.~\ref{fig:rib}. However, true textiles are subject to inter-yarn friction and variable yarn tension, each of which alters the relaxed configuration of a stitch as well as its mechanical response.
For example, the process of knitting involves looping pre-tensioned yarn over needles or latch-hooks, imparting stress on `downward-facing' arcs in each structure.
While the resulting deformation preserves mirror symmetries, there is a loss of two-fold rotation and inversion symmetries.

Patterns of general $m \times n$ garter, rib and seed are also knittable. For example, $2 \times 1$ garter would have two rows of knit followed by one row of purl, repeated in the vertical direction and $2 \times 2$ seed has a four by four-square checkerboard of knits and purls as its unit cell. These fabrics' symmetries are distinct from those of their fundamental counterparts owing to the additional symmetries introduced by replicating pieces of the fundamental unit cell. Correspondingly, the mechanical response of an $m \times n$ pattern will be distinct from that of its fundamental patterns. In our discussions, we have restricted to the variations of garter, rib, and seed with $m = n = 1$.  We leave the cases of $ m \neq 1$ or $n \neq 1$ to future investigations.

Finally, we discuss implications of these bulk symmetries on the boundaries of the fabric. 
In rib and seed stitches, mirror symmetry is broken along half of the mirror planes for which it existed in garter and stockinette stitches, and this instead becomes a two-fold rotation axis pointing along the wale direction. 
The direction of fabric loops along these two-fold rotation axes points to the direction that the fabric was cast-on. 
Thus, by examining only a single unit cell of rib or seed fabric one can determine its direction of origin, a distinction impossible in garter or stockinette fabric.

\section{Conclusion}

In this article we extend the classical Conway notation from the planar wallpaper groups to the three-dimensional setting of the thickened plane. By introducing new symbols for the singular features unique to layer-group actions and establishing a cost formalism that generalizes Conway’s Magic Theorem, we obtained a complete, coordinate-independent, and topologically meaningful classification. This provides a compact description of the quotient spaces associated with layer-group symmetries and complements the traditional Hermann–Mauguin notation. To illustrate the applicability of this framework, we analyzed the symmetry of common knitted patterns, showing how their layer-group symmetries are naturally captured by the new orbifold symbols.  

Recent work~\cite{Singal2023} has shown that the internal, sub-unit cell symmetries of knitted fabrics can rationalize the macroscopic response of fabrics, expanding on previously-developed geometric models~\cite{Postle1967_a,Postle1967_b,Poincloux2018} of knitted fabric mechanics to include a larger set of patterns.
In particular, in a reduced elastica model of 2D projections of the full 3D yarn shape, it was found that curve segments exhibiting `even symmetry'---associated with mirror symmetries---are stiffer than segments exhibiting `odd symmetry'---associated with inversion symmetries or 2-fold axes---by an order of magnitude. 
Since the curve segments possessing these symmetries are additionally either primarily oriented along the course or the wale directions, their responses to applied loading conditions are highly directional.
This combination of symmetry and yarn orientation was shown~\cite{Singal2023} to rationalize the observed pronounced anisotropic elastic response of the different knit patterns as well as reasonable predictions for the comparative magnitude of fabric stiffnesses.
We also here note related work that simulated the behavior of the four basic stitches using a yarn-based dynamical model~\cite{Ding2024}. The binary decomposition of curves into even and odd shapes used in the analysis of~\cite{Singal2023} is only applicable in a 2D approximation of the curve shapes. 
Our analysis of knitted structures using full layer symmetries is the first step towards a potential generalization of this prior work, where distinct responses for curve segments with local mirror, inversion, and rotational point group symmetries could be used to construct a more complete coarse-grained mechanical model of knitted fabrics. This would complement additional recent work on elasticity and other tensorial properties of anisotropic doubly periodic materials which were determined using representation theorems and shown to be highly dependent on the wallpaper group of the material~\cite{Dresselhaus2025AnomalousTensorial2D}.

The use of symmetry to rationalize the mechanical response of knitted fabrics is a clear over-simplification of a material dependent on the complex interplay of packing and friction within and between spun yarn, as well as preparation and wear history~\cite{hearle1969structural}.
True fabrics do not possess the idealized symmetries discussed here; instead, owning to fiction, variation in yarn properties, and variations in tension, fabric structure can be considerably perturbed from the ideal, potentially with both spatially correlated and un-correlated randomness.
Nevertheless, condensed matter is replete with examples of disordered materials without \textit{global symmetries} whose electronic~\cite{wearie,spring2021_amorphous_topological_phases,fulga2014_statistical_topological_insulators} and mechanical~\cite{mitchell2018_amorphous_topological_insulators} properties depend the \textit{local symmetries} that can be identified in material microstructure.
We thus hypothesize that the symmetry classification of knitted fabrics, and the mechanical implications thereof, may apply to fabrics with inevitable imperfections.
Future explorations will also require the creation of an algorithm to facilitate identification of layer group symmetries, building on the approach for identifying wallpaper symmetries~\cite{sym_of_things}, as well as automated approaches for identifying `approximate symmetries' in real layer materials~\cite{Gandhi2021}.

In conclusion, our work demonstrates how the extended orbifold notation developed here can support the study of doubly periodic structures in materials science, where geometry, connectivity, and symmetry jointly influence material behaviour. 
The framework developed here opens the way for future investigations, including the systematic classification of woven and knitted textiles as well as chainmail~\cite{Farris2020,Farris2021,Klotz2024}, origami~\cite{lv2014_origami_metamaterials}, and the analysis of periodic entanglement in engineered materials~\cite{moestopo2023_knots_microarchitected_materials, pescialli2025_topology_informed_architected}. It may also apply to entangled polymer networks~\cite{kim2021_fracture_polymers, dhand2024_additive_entangled_polymers}, bio-inspired entangled materials~\cite{fox2024_orange_pericarps} and broader doubly periodic structures arising in mathematics and physics.

\section*{Acknowledgements}

This article is dedicated to Stephen T. Hyde, whose influential contributions to the study of orbifold symmetry in triply periodic entangled structures and their applications in chemistry were an important source of inspiration for this work.

The authors are grateful to the anonymous referees for their careful reading and valuable comments, which helped improve the clarity and quality of the paper. In particular, we thank them for pointing out an error in the analysis of the underlying surfaces of the orbifolds and for guiding us toward the more cautious treatment.

The authors also thank Elisabetta Matsumoto for valuable discussions regarding the symmetry classification of knitted fabrics.

The first author is supported by JSPS KAKENHI Grant-in-Aid for Early-Career Scientists (Grant Number 25K17246) and the Daiichi-Sankyo ``Habataku'' Support Program for the Next Generation of Researchers 2025. She also acknowledges financial support from the Tohoku Forum for Creativity of Tohoku University, the Japan Tourism Agency (MICE), and the Sendai Tourism, Convention and International Association (SenTIA) for supporting the program \textit{The Theory of Periodic Tangles and Their Interdisciplinary Applications}, during which a significant part of this research was developed.
The second author acknowledges support from the National Science Foundation (Grant No.\ CMMI-2344589), the College of Chemistry at UC Berkeley, and the support of Sanjay Govindjee and Kranthi Mandadapu.

\section*{Competing interests} 
We declare we have no competing interests.

\bibliography{library}

\end{document}